\documentclass[article]{aa}

\usepackage{natbib}
\usepackage{graphicx}
\usepackage{txfonts}

\begin{document}

\title{The role of Rayleigh-Taylor instabilities in filament threads} \author{}
\author{J. Terradas, R. Oliver \& J. L. Ballester}

\offprints{J. Terradas, \email{jaume.terradas@uib.es}}
\institute{Departament de F\'\i sica,
Universitat de les Illes Balears, E-07122, Palma de Mallorca, Spain}
{}

\date{Received / Accepted }

\abstract{Many solar filaments and prominences show short-lived horizontal
threads lying parallel to the photosphere.} {In this work the possible link
between Rayleigh-Taylor instabilities and thread lifetimes is investigated.}
{This is done by calculating the eigenmodes of a thread modelled as a Cartesian
slab under the presence of gravity. An analytical dispersion relation is derived
using the incompressible assumption for the magnetohydrodynamic (MHD)
perturbations.} {The system allows a mode that is always stable, independently
of the value of the Alfv\'en speed in the thread. The character of this mode
varies from being localised at the upper interface of the slab when the magnetic
field is weak, to having a global nature and resembling the transverse kink mode
when the magnetic field is strong. On the contrary, the slab model permits
another mode that is unstable and localised at the lower interface when the
magnetic field is weak. The growth rates of this mode can be very short, of the
order of minutes for typical thread conditions. This Rayleigh-Taylor unstable
mode becomes stable when the magnetic field is increased, and in the limit of
strong magnetic field it is essentially a sausage magnetic mode.}{The gravity
force might have a strong effect on the modes of oscillation of threads,
depending on the value of the Alfv\'en speed. In the case of threads in
quiescent filaments, where the Alfv\'en speed is presumably low, very short
lifetimes are expected according to the slab model. In active region
prominences, the stabilising effect of the magnetic tension might be enough to
suppress the Rayleigh-Taylor instability for a wide range of wavelengths.}


\titlerunning{Rayleigh-Taylor instabilities in threads}
\authorrunning{Terradas et al.}
\maketitle
\newcommand{\etal}{{et al.}}

\section{Introduction}

Threads are thought to be the building blocks of solar filaments and prominences.
These fine-scale structures are readily seen in high quality images obtained using
the Swedish Solar Telescope (SST), the Dutch Open Telescope (DOT), or the Solar
Optical Telescope (SOT) onboard {\em Hinode}. Threads are a part of a magnetic tube
filled with cold plasma with densities much larger than the coronal surrounding,
the density contrast being of the order $100$. Their observed widths are typically
between $100\, \rm km$ and $600\, \rm km$, while their lengths are in the range
$3500-28000\, \rm km$ \citep[see][]{linetal05}. The total length of the magnetic tube where the thread is
supposed to be embedded is much longer than the thread length, and in many cases is
of the order of $10^5\, \rm km$. Some threads are essentially horizontal, i.e.,
running parallel to the photosphere, and are usually located along the spine of the
filament \citep[see][for a clear example]{okamoto07}. Inclined threads are mostly
found in filament barbs, whereas vertical threads are clearly observed in hedgerow
prominences \citep[see for example][]{bergeretal08,chaeetal08}.

Threads have short lifetimes, and in quiescent filaments they are in the range
$9-20\, \rm min$ \citep[see the review of][]{lin11}. The mechanisms that produce
the thread disappearance are usually thought to be related to mass flows and
thermal instabilities \citep[see for example,][in the context of
prominences]{carbetal04,soleretal11a,soleretal12}. However, other types of
instabilities may play a role. Recently, \cite{ryuetal10} have shown that a
number of processes taking place in prominences during their evolution can be
linked to fundamental fluid instabilities such as Rayleigh-Taylor and
Kelvin-Helmholtz instabilities \citep[see also][]{bergeretal10}. Using a single interface model between the
prominence and corona, and the assumption of incompressibility for the
perturbations, these authors have been able to attribute the appearance of
plumes, spikes and ripples to these instabilities. \cite{hillieretal11,hillieretal12} have
numerically investigated the nonlinear stability of the  Kippenhahn-Shl\"uter
prominence model to the Rayleigh-Taylor instability.

An additional feature of threads is that they show oscillations. For example,
\cite{thomsch91,yieng91,yietal91} found that some threads seem to oscillate
independently while others appear to oscillate together \citep[see the reviews
of][]{oliball02,baneretal07,arreguietal12}. More recent high spatial resolution and high
cadence observations have revealed the presence of both propagating and standing
waves in individual threads \citep[see][]{linetal07,linetal09,okamoto07}. For
example, \citet{linetal07} found oscillations with wavelengths of the order of  
$3000\, \rm km$ and periods around $5.4\, \rm min$ in quiescent filaments, while
\citet{linetal09} have reported periods of 3.6 minutes in swaying filament
threads. \citet{okamoto07} have detected waves in an active region prominence
with wavelengths of at least $2\times 10^5\, \rm km$ with periods of $4\,\rm
min$. Note that the periods of oscillation are typically of the same order of
magnitude as the lifetimes of threads in quiescent filaments.

The theoretical interpretation of the reported oscillations in threads is
usually done by means of magnetohydrodynamic (MHD) waves. A  certain prominence
model is assumed and the eigenmodes of oscillation are determined
\citep[see][for a detailed description of the results using different
configurations]{oliball02,arreguietal12}.  In particular, two basic models have
been considered in the literature, the Cartesian slab
\citep[see][]{joarderetal97, diazetal01,diazetal05}, and the cylindrical tube
\citep[see for
example,][]{diazetal02,dymrud05,soler10,arreguiterretal08,arreguietal11}. MHD
waves in cylindrical plasmas are characterised by two wavenumbers, the axial
wavenumber and the azimuthal wavenumber. For typical threads the azimuthal
wavenumber of the detected MHD waves is much larger than the longitudinal
wavenumber. The advantage of the cylindrical tube is that it is more realistic
than the Cartesian slab which is, in general, two dimensional and thus infinite
in the third direction. However, the inclusion in the slab configuration of
propagation in this third direction, equivalent to the azimuthal component in a
cylindrical tube, mimics the effect of a three-dimensional structure regarding
the MHD waves.   

The motivation of the present work is to study oscillations of horizontal
threads under the presence of the gravitational force. This problem has been
partially addressed by \cite{mcewandiaz07} using a different approach. Here the
thread is modelled using the slab configuration. The incompressible assumption
is adopted for the perturbations which are allowed to be three-dimensional. The
eigenmodes of the system are carefully analysed, paying special attention to the
regime where the system becomes unstable due to the Rayleigh-Taylor instability,
driven by the gravity force. Our focus is on the possible relation between this
instability and the short lifetimes of threads.

\section{Single interface results}\label{inteface}

\begin{figure}[!hh]
\center{\includegraphics[width=8.5cm]{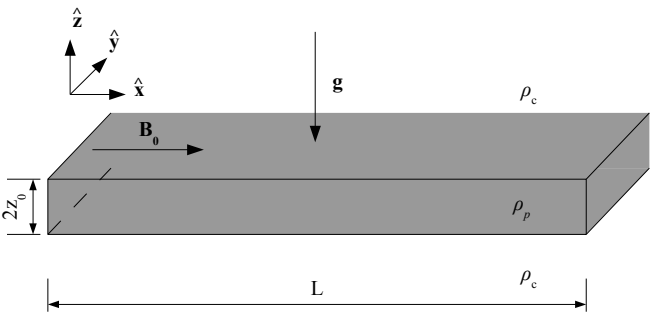}}
 \caption{\small Sketch of the slab model representing an idealised thread.
}\label{fibril}
\end{figure}

It is well known that a single interface between fluids with different
densities, like the interface between threads and the surrounding corona,  can 
be unstable in the presence of gravity when the upper fluid is heavier that the
lower fluid, this is the Rayleigh-Taylor instability
\citep{rayleigh1883,taylor50}. Let us assume that we have an interface in the
$z-$direction. A constant magnetic field, ${\bf B_0}$, is imposed to be along the 
$x-$direction, and gravity, ${\bf g}$, is acting across the magnetic field. The
sketch of  the configuration is represented in Figure \ref{fibril} for a double
interface, but at this point we just consider a single interface. We assume that
the equilibrium density, $\rho_0$, is constant at each side of the interface
($\rho_0=\rho_{\rm p}$ for $z>0$, while $\rho_0=\rho_{\rm c}$ for $z<0$). Since the density is a piecewise constant magnitude the equilibrium gas pressure,
$p_0$,  has to change linearly with $z$ due to the presence of the gravity
force, and must be continuous at the interface to have hydrostatic balance.

We use the incompressible linearised ideal MHD equations for the perturbations.
We seek for solutions with a dependence on $t$, $x$ and $y$ of the form
$e^{i\omega t + ik_x x +i k_y y}$, where $\omega$ is the frequency and  $k_x$
and $k_y$ are the wavenumbers parallel and perpendicular to the magnetic
field, respectively. The total wavenumber is $k=\sqrt{k_x^2+k_y^2}$. The
incompressibility constraint is
\begin{eqnarray}\label{incompress}
\nabla \cdot {\bf v}=i k_x v_x+ i k_y v_y+ \frac{dv_z}{dz}=0.
\end{eqnarray}
Under such assumption the continuity equation for the
perturbed density reduces, in our configuration, to
\begin{eqnarray}\label{cont} i \omega \rho=-v_z \frac{d\rho_0}{dz}.
\end{eqnarray}
From the momentum equation we have
\begin{eqnarray}\label{mom}
i \omega\rho_0v_x&=&-i k_x p, \label{vx}\\
i \omega\rho_0v_y&=&-i k_y p+\frac{B_0}{\mu}\left(i k_x b_y-i k_y b_x\right), \label{vy}\\
i \omega\rho_0v_z&=&-\frac{dp}{dz}+\frac{B_0}{\mu}\left(i k_x
b_z-\frac{db_x}{dz}\right)-\rho g, \label{vz}
\end{eqnarray}
while from the induction equation we obtain
\begin{eqnarray}
i \omega b_x&=&B_0 i k_x v_x, \label{bx}\\
i \omega b_y&=&B_0 i k_x v_y, \label{by}\\
i \omega b_y&=&B_0 i k_x v_z. \label{bz}
\end{eqnarray}
Note that the equilibrium gas
pressure, $p_0$, does not appear in the previous equations because of the incompressible
assumption, but the perturbed pressure, $p$, is in the equations although we do not
have an explicit equation for this magnitude. 

We want to obtain a single equation for
$v_z$, and the first step is to insert the corresponding components of the perturbed magnetic
field given by Eqs.~(\ref{bx})-(\ref{bz}) in Eqs.~(\ref{vy})-(\ref{vz}) and make
use of Eq.~(\ref{cont}). The following equations are obtained
\begin{eqnarray}
i \omega\rho_0v_y&=&-i k_y p+ \frac{B_0^2}{\mu}\frac{k_x}{\omega}\left(i k_x
v_y-i k_y v_x\right), \label{vym}\\
i \omega\rho_0 v_z&=&-\frac{dp}{dz}+\frac{B_0^2}{\mu}\frac{k_x}{\omega}  \left(i
k_x v_z-\frac{dv_x}{dz}\right)+\frac{1}{i \omega} g \frac{d\rho_0}{dz} v_z. \label{vzm}
\end{eqnarray}
Eliminating the pressure perturbation from Eqs.~(\ref{vx}) and (\ref{vym}) we
find that 
\begin{eqnarray}
i k_x
v_y-i k_y v_x=0,\label{vort}
\end{eqnarray}
thus, the second term in the right hand side of Eq.~(\ref{vym}) is zero.
Using Eqs.~(\ref{vx}) and (\ref{vym}) together with Eq.~(\ref{incompress}) we find the following expression for the
perturbed gas pressure
\begin{eqnarray}\label{press}
p=-i \rho_0\frac{\omega}{k^2} \frac{dv_z}{dz}.
\end{eqnarray}
In addition, combining Eqs.~(\ref{incompress}) and (\ref{vort}) we obtain
\begin{eqnarray}\label{vxs}
v_x=i \frac{k_x}{k^2} \frac{dv_z}{dz}.
\end{eqnarray}
\noindent Now inserting Eqs.~(\ref{press}) and (\ref{vxs}) in Eq.~(\ref{vzm}) we find the
final equation for $v_z$
\begin{eqnarray}
\frac{d}{dz}\left(\rho_0 \frac{dv_z}{dz} \right)+\frac{B_0^2}{\mu}
\frac{k_x^2}{\omega^2} \left(k^2-\frac{d^2}{dz^2}\right) v_z-k^2 \left(\rho_0+
\frac{g}{\omega^2}\frac{d\rho_0}{dz}\right) v_z=0.\nonumber \\
 \label{vzmf}
\end{eqnarray}
It is easy to see that for a piecewise density profile the
eigenfunctions  have the following
simple form 
\begin{eqnarray}\label{eigenfuninter}
v_z(z)=\left\{
\begin{array}{ll}
A\, e^{kz}, & \; z< 0, \\
B\, e^{-kz}, & \; z> 0.
\end{array}
\right.
\end{eqnarray}
The constants $A$ and $B$ are calculated requiring continuity of $v_z$ across
the interface (meaning that $A=B$). There is an additional condition that comes from
the integration of Eq.~(\ref{vzmf}) over an infinitesimal element of $z$ that
includes the interface (located at $z=0$). This condition reduces to 
\begin{eqnarray}\label{jumpcond}
\left(-\omega^2\rho_{\rm p} + k_x^2\frac{B_0^2}{\mu}\right)\left(\frac{d
v_z}{dz}\right)^++\left(\omega^2\rho_{\rm c} - k_x^2\frac{B_0^2}{\mu}\right)\left(\frac{d
v_z}{dz}\right)^- \nonumber \\
=-g k^2 \left(\rho_{\rm p}-\rho_{\rm c}\right) v_z(z=0),
\end{eqnarray}
where we have used that $\rho_0=\rho_{\rm p}$ for $z>0$ (``+" region) and
$\rho_0=\rho_{\rm c}$  for $z<0$ (``-" region). Inserting the derivatives of
$v_z$ at each side of the interface in Eq.~(\ref{jumpcond}) leads to the well known dispersion relation for the incompressible
interface problem \citep[see also][]{chandra61}
\begin{eqnarray}\label{instinter}
\omega^2=-g k \frac{\rho_{\rm p}-\rho_{\rm c}}{\rho_{\rm p}+\rho_{\rm c}}+k_x^2
\frac{2 B_0^2}{\mu\left(\rho_{\rm p}+\rho_{\rm c}\right)}.
\end{eqnarray}
\noindent 
From this equation it is clear that in the absence of magnetic field the
system is always unstable ($\omega^2<0$) if $\rho_{\rm p}>\rho_{\rm c}$, i.e., when the
heavier fluid is on top of the lighter fluid. The opposite configuration is
always stable, i.e., $\omega^2>0$ for $\rho_{\rm p}<\rho_{\rm c}$. The unstable mode can become stable if the second term of the
right-hand side of Eq.~(\ref{instinter}) is larger than the absolute value of
the first term. This term is simply the square of the kink frequency.  It is interesting to note that
Eq.~(\ref{instinter}) simplifies to 
\begin{eqnarray}\label{instintersim}
\omega^2=-g k +k_x^2
\frac{2 B_0^2}{\mu \rho_{\rm p}},
\end{eqnarray}
if the density at one side of the interface is much larger that the density at the
other side ($\rho_{\rm p}\gg \rho_{\rm c}$, in the previous equation). This is a common
characteristic in prominence threads.

\section{Slab model results}\label{slab}  

The interface model is useful to understand the basics of the Rayleigh-Taylor
instability. However, the main purpose of this work is to study threads, and
these structures are better represented using slab or cylindrical models. Here
we assume a slab configuration. Across the field, i.e., in the $z-$direction, we
assume again that density is constant. The  equilibrium gas pressure is chosen
to have hydrostatic equilibrium, and as in the interface model it changes
linearly with $z$. The density takes the value $\rho_{\rm p}$ inside the slab,
representing a thread located between $-z_0$ and $z_0$, and $\rho_{\rm c}$
outside the slab (see Figure \ref{fibril}). In this configuration the Alfv\'en
speed has a constant value inside and outside the thread. Notice that the thread
is infinite in the $y-$direction. The model adopted here is similar to that
studied in \cite{mcewandiaz07}, where the main distinctive difference 
is the incompressible assumption for the MHD waves and specially the treatment
of the boundary conditions at the interfaces, given in our case by
Eq.~(\ref{jumpcond}). \citet{goedbloedpoedts04} analysed a gravitating slab
supported from below by a vacuum magnetic field but conducting wall boundary
conditions were considered \citep[see also the works
of][]{hermansgoossens1987,hermansgoossens1989}.

The slab configuration is essentially the combination of two interfaces.
Extending the single interface eigenfunction, given by
Eq.~(\ref{eigenfuninter}),  to the double interface problem, leads to 
\begin{eqnarray}\label{eigenslab}
v_z(z)=\left\{
\begin{array}{lll}
A\, e^{k\,(z+z_0)}, & \; z<-z_0,\\
B\, e^{-k\,(z+z_0)}+C\, e^{k(z-z_0)},
& \; z\leq|z_0|, \\
D\, e^{-k\,(z-z_0)}, & \; z> z_0.
\end{array}
\right.
\end{eqnarray}
Continuity of $v_z$ at $z=\pm z_0$ imposes a relation between constants
\begin{eqnarray}\label{constslab}
B&=&A-C\,e^{-2 k z_0},\\
D&=&C+B\,e^{-2 k z_0}.
\end{eqnarray}

\noindent Since one of these constants can be chosen arbitrarily we assume that
$A=1$. The constant $D$ is written, using the equation for $B$, as a function of
the constant $C$ only.  The next step is to apply the second boundary condition, given
by Eq.~(\ref{jumpcond}),
at each interface ($z=\pm z_0$) using the appropriate values for the densities. These boundary conditions
provide two equations for $\omega^2$ in terms of the constant $C$ which is
still unknown. The explicit form of this is calculated by combining
these two equations. 

There are two solutions for $C$ representing two different
types of modes, hereafter denoted by the subscripts ``$+$" and ``$-$". The final dispersion
relation is the following algebraic equation
\begin{eqnarray}\label{instslab} \omega^2=\frac{-g k \left(\rho_{\rm
p}-\rho_{\rm c}\right)+k_x^2 \frac{2 B_0^2}{\mu} \left(1- C_\pm e^{-2 k
z_0}\right)}{\rho_{\rm p}\left(1-2 C_\pm e^{-2 k z_0}\right)+\rho_{\rm
c}},
\end{eqnarray}

where
\begin{eqnarray}\label{ceq}
C_\pm=\frac{E \pm F}{G},
\end{eqnarray}
being
\begin{eqnarray}\label{eeq}
E=e^{2 k {z_0}} g k
\left(1+\frac{\rho_{\rm c}}{\rho_{\rm p}}\right) \sinh{2 k {z_0}}-k_x^2 v_{\rm Ap}^2,
\end{eqnarray}
\begin{eqnarray}\label{feq}
F&=&e^{2 k {z_0}} \left \{ \left({k_x^2
v_{\rm Ap}^2}\right)^2\right.\nonumber \\ & & \left.
+g^2 k^2 \left[\left(1+\frac{\rho_{\rm c}^2}{\rho_{\rm p}^2}\right) \sinh^2{2 k
{z_0}}+\frac{\rho_{\rm c}}{\rho_{\rm p}} \sinh{4 k {z_0}}\right]\right\}^{1/2},
\end{eqnarray}
and
\begin{eqnarray}\label{geq}
G=2 \sinh{2 k {z_0}} \left(k_x^2 v_{\rm Ap}^2+g k
\right).
\end{eqnarray}
In the previous expressions we have introduced the Alfv\'en speed in the thread,
$v_{\rm Ap}=B_0/\sqrt{\mu \rho_{\rm p}}$. It is clear that the slab dispersion relation given by Eq.~(\ref{instslab}) has
some similarities with the interface dispersion relation
(Eq.~(\ref{instinter})), but it is more involved due to the dependence of $C$ on
the different parameters. For a better understanding of the solutions some
limiting cases are investigated in the following subsections.

Before we analyse the different cases we need to discuss the values of some of the
parameters that appear in the dispersion relation. Although we are considering an
infinite slab model in the $y-$direction we incorporate the three-dimensionality of
the motion by choosing the appropriate $k_y$. This is an important parameter in our
model. If we suppose that the thread is a magnetic tube of circular cross-section
then $k_y=m/R$ where $m$ is the azimuthal wavenumber and $R$ is the radius of the
thread. For transverse motions $m=1$ (since the system is periodic in the azimuthal
direction $m$ is restricted to be an integer). The equivalent mode in the slab has
$k_y=1/z_0$ ($z_0$ in the slab is equivalent to $R$ in the cylinder) and
corresponds to a transverse motion of the tube if $k_x\ll k_y$. The longitudinal
wavenumber is $k_x=2 \pi/\lambda$, being $\lambda$ the characteristic wavelength.
It is clear that $k_x/k_y$ is of order $R/\lambda$. The typical radius of threads
is $R \sim 10^2 \, \rm km$, while the reported wavelengths are in the range
$\lambda=10^3-10^5 \, \rm km$. This means that the regime of interest is when
$R/\lambda\ll 1$, i.e. in the thin tube approximation.

\subsection{Purely magnetic case}\label{magnsec}
 
\begin{figure}[!hh] \center{\includegraphics[width=8cm]{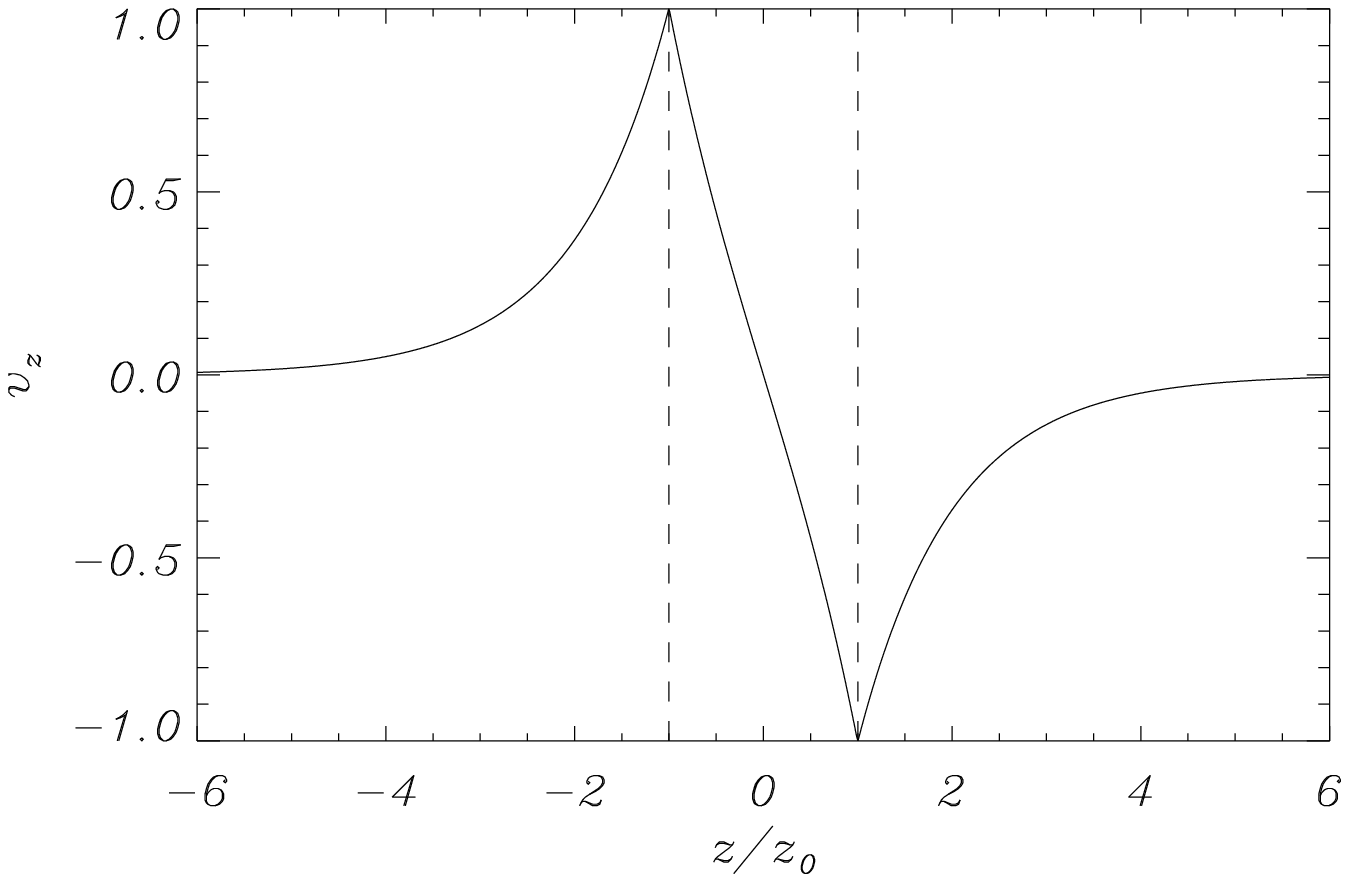}}
\center{\includegraphics[width=8cm]{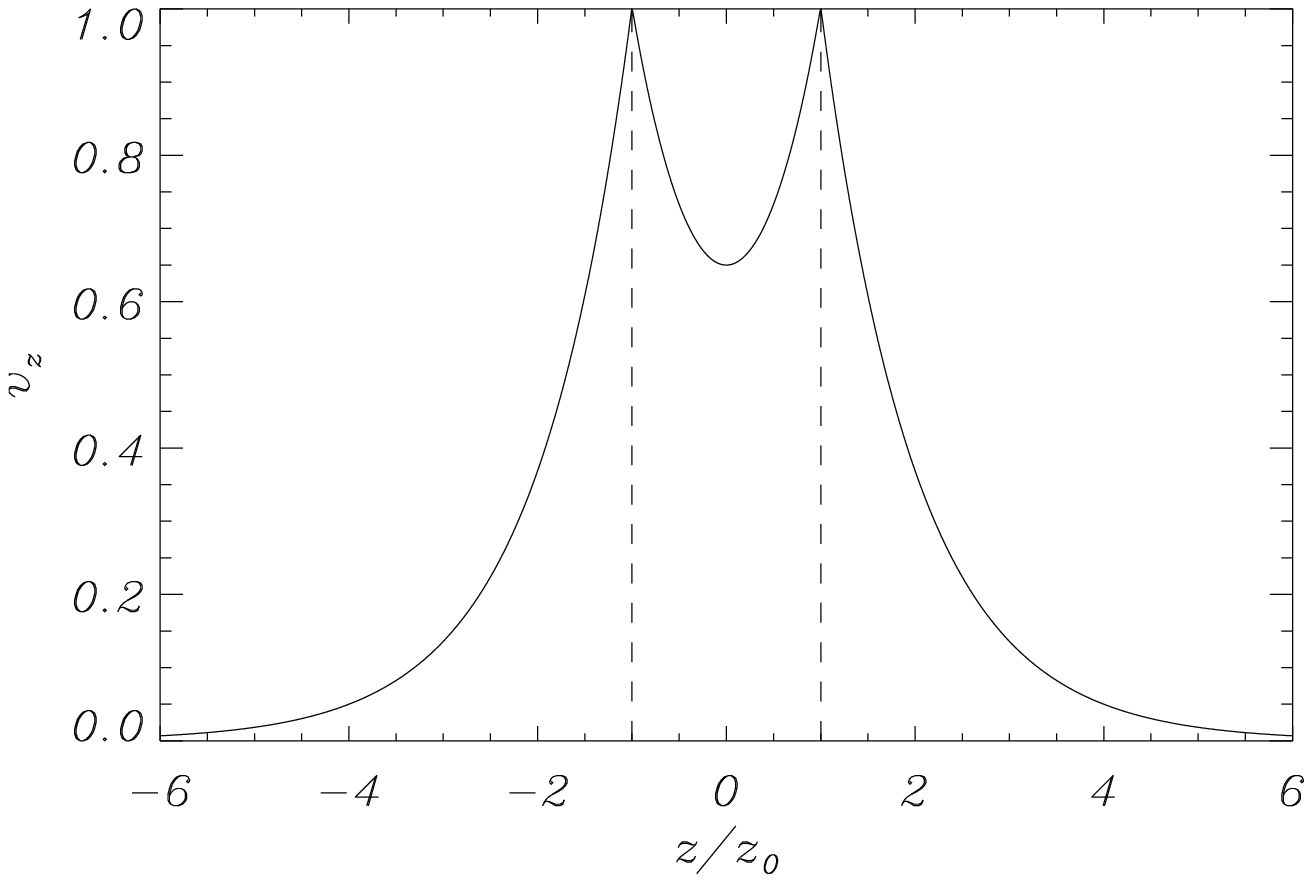}}  \caption{\small Sausage 
(top panel) and
kink 
(bottom panel) eigenfunctions in the purely magnetic case. In this plot $R=10^2 \, \rm
km$,  $\lambda=2\times 10^5 \, \rm km$, $\rho_{\rm p}=100 \rho_{\rm c}$, $v_{\rm
Ap}=100\,\rm km\, s^{-1}$ and $g=0$. The dashed lines represent the slab
boundaries. }\label{vzeigeng0} \end{figure}

We consider first the case when gravity is absent and we analyse the properties
of the eigenmodes. The system has symmetric or antisymmetric eigenfunctions,
meaning that either $B=C$ (for the ``$+$" solution) or $B=-C$ (for the ``$-$"
solution), and the eigenfunctions inside the slab  are the 
hyperbolic functions ``cosh" or ``sinh", respectively. The symmetric  mode,
or kink, satisfies that $D=A$, while for the  antisymmetric or sausage mode
$D=-A$. In Figure \ref{vzeigeng0} the eigenfunctions are plotted for a
particular choice of parameters representative of thread conditions,  $R =
10^2\, \rm km$ ($k_y=1/R$), $\lambda=2\times 10^5 \, \rm km$
($k_x=2\pi/\lambda$),  $\rho_{\rm p}=100 \rho_{\rm c}$, and $v_{\rm Ap}=100\,\rm
km\, s^{-1}$. Contrary to the eigenmodes in a cylindrical tube
\citep[see][]{spruit82,edrob83,cally86,cally03,goossensetal09}, the sausage mode
in the slab model is trapped, instead of leaky. Additionally, the eigenfunction
of the kink mode has a minimum at the centre of the slab instead of a maximum,
but since the mode essentially produces a displacement of the whole slab, we
sill refer to this mode as the transverse mode \citep[see also][for the spatial
distribution of the eigenfunctions in the compressional problem]{arrterretal07}.

It can be shown from Eqs.~(\ref{ceq})-(\ref{geq}) that in the case of zero
gravity the constants are $C_\pm=1/\left(e^{-2 k {z_0}}\pm 1\right)$, and  the
modes are always stable. The dispersion relation without gravity reduces to    
\begin{eqnarray}\label{disperhomslabk}
\omega^2=k_x^2\frac{B_0^2}{\mu}\frac{1+\coth(k z_0)}{\rho_{\rm p}+\rho_{\rm c}
\coth(k z_0)}, \end{eqnarray} for the kink mode and  
\begin{eqnarray}\label{disperhomslabs}
\omega^2=k_x^2\frac{B_0^2}{\mu}\frac{1+\tanh(k z_0)}{\rho_{\rm p}+\rho_{\rm c}
\tanh(k z_0)}, \end{eqnarray}

\noindent for the sausage mode. These equations agree with the results of
\citet{edrob82} when $k_y=0$ and in the incompressible limit (see their
Eq.~(12)). When $k z_0$ is large the hyperbolic tangent and cotangent tend to
one, and we recover the dispersion relation for a single interface (see
Eq.~(\ref{instinter})) when gravity is absent, i.e., the frequency tends to the
kink frequency. For a finite slab and when $k_x\ll k_y$ the frequency of these
modes tends again to the kink frequency  \citep[this is true even for the
compressible case, see][]{arrterretal07}, and this is the regime we are
interested in.

\subsection{Purely gravitational case}

We now analyse the situation when gravity is present and $B_0=0$. 
For this case it can be shown that Eq.~(\ref{instslab}) simplifies to the following
dispersion relation
\begin{eqnarray}\label{nofield} 
\omega^2=\pm \frac{g k
\left(\rho_{\rm p}-\rho_{\rm c}\right) \sinh{2 k
z_0}}{\sqrt{\left(\rho_{\rm p}^2+\rho_{\rm c}^2\right)\sinh^2{2 k
z_0}+\rho_{\rm p} \rho_{\rm c} \sinh{4 k
z_0}}}.
 \end{eqnarray}

\begin{figure}[!hh] \center{\includegraphics[width=8cm]{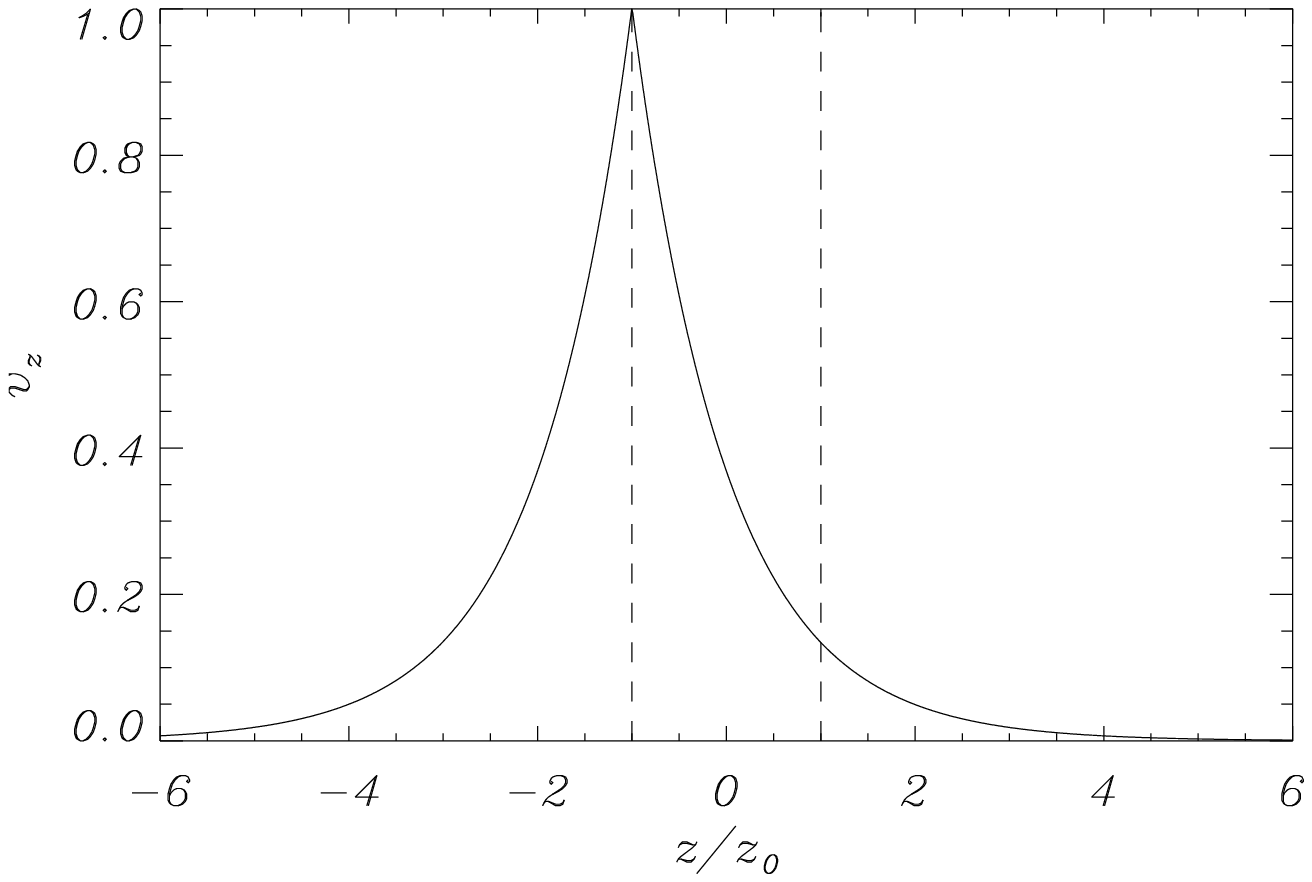}}
\center{\includegraphics[width=8cm]{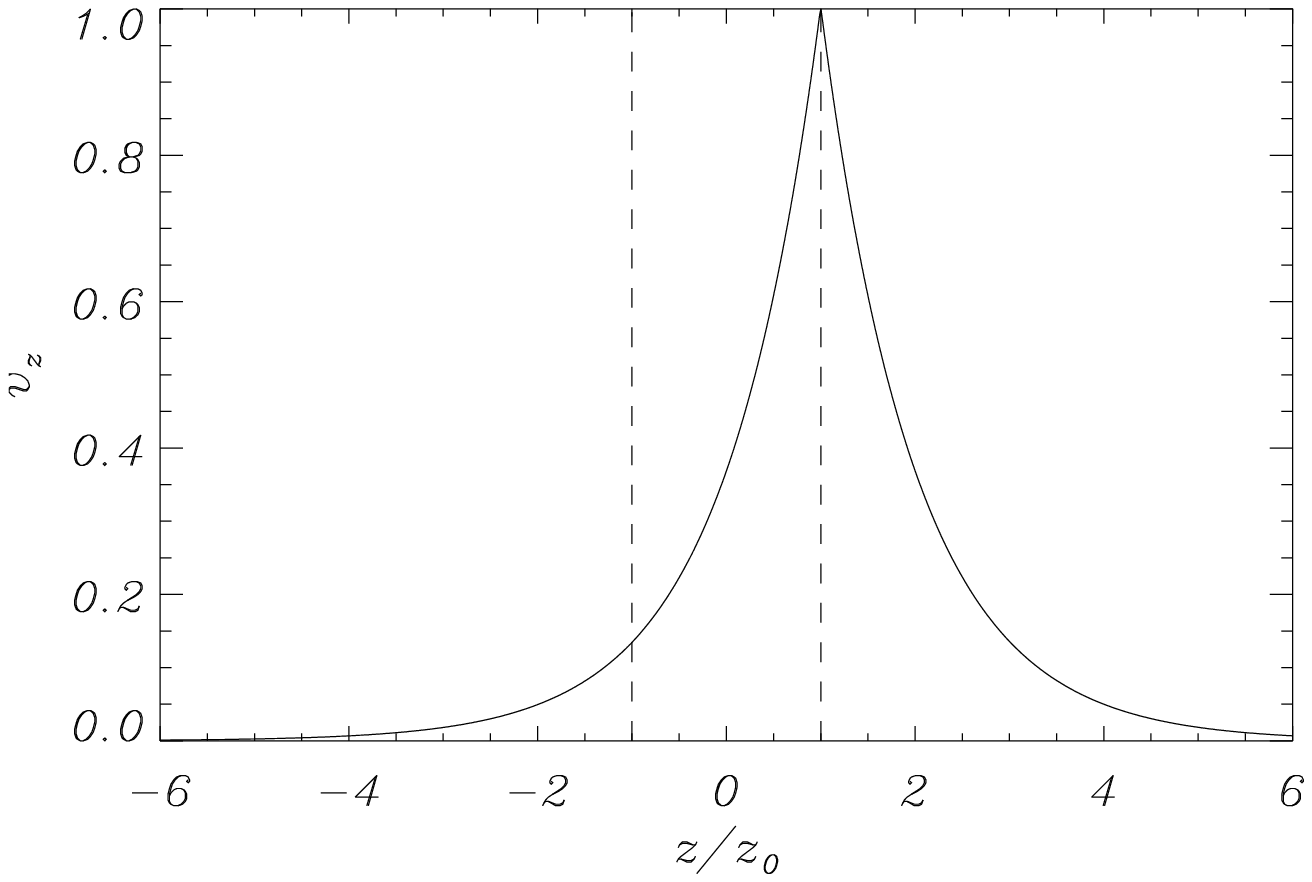}}  \caption{\small Eigenfunction
of the modes associated to the lower unstable mode  (top panel) and upper
stable mode (bottom panel) in the purely gravitational case. In this plot
$R=10^2 \, \rm km$, $\lambda=2\times 10^5 \, \rm km$, $\rho_{\rm p}=100
\rho_{\rm c}$ and $v_{\rm Ap}=0$ ($g=0.274\, \rm  km\, s^{-2}$). The dashed
lines represent the slab boundaries. }\label{vzeigenb0} \end{figure}

\noindent This configuration has a stable mode always associated to $C_+$, and
an unstable solution linked to the $C_-$ constant. The single non-magnetic
interface dispersion relation, given by Eq.~(\ref{instintersim}) without the
magnetic term, is easily recovered from Eq.~(\ref{nofield}) when $\rho_{\rm
p}\gg \rho_{\rm c}$ (the second term in the denominator can be neglected in
front of the first term). If $k z_0\gg 1$  the dispersion relation tends also to
the interface solution since the wavelength of the perturbation is much smaller
than the half width of the slab. The spatial distribution of the corresponding
eigenfunctions is rather different respect to the kink and sausage magnetic
eigenfunctions. It is possible to see that the stable mode is mainly localised
at the upper interface since for this mode $C\simeq D$ ($C_+$ tends to $e^{-2 k
{z_0}}$ according to Eqs.~(\ref{ceq})-(\ref{geq})),  while the unstable mode
satisfies that $B\simeq A$ ($C_-$ tends to 0), and it is therefore located at
the lower interface (in this case the heavier fluid is on top of the lighter
fluid and for this reason the mode is unstable). An example of the eigenfunctions is found in Figure \ref{vzeigenb0}. The
degree of localisation of the eigenfunction around the interfaces is given by
the penetration length, which according to Eq.~(\ref{eigenslab}), can be defined
as $l=1/k$. For the situation $k_x\ll k_y$  and $k_y=1/z_0$ we have that
$l\approx z_0$ ($k\approx k_y$). Hence,  for these parameters, slab eigenmodes
due to gravity do not represent a coherent motion of the whole slab like in the
kink mode, and are mostly surface waves associated to the individual interfaces.

\subsection{Full case} 

\begin{figure}[!hh] \center{\includegraphics[width=8cm]{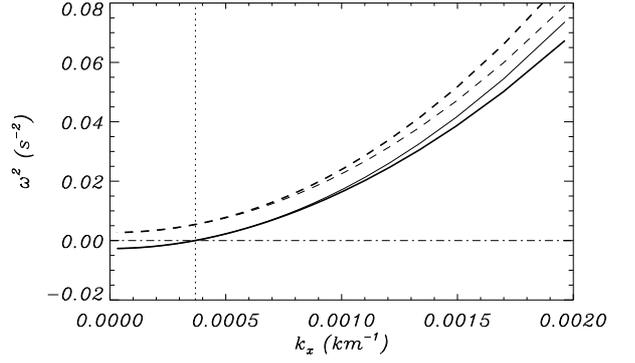}} 
\caption{\small Square of the frequency as a function of $k_x$. In this plot
$R=10^2 \, \rm km$, $\rho_{\rm p}=100 \rho_{\rm c}$, and $v_{\rm Ap}=100\,\rm km\, s^{-1}$. Thick lines correspond
to the full dispersion relation for the slab (Eq.~(\ref{instslab})), while thin
lines represent the solutions of the single interface (Eq.~(\ref{instinter})).
 The dashed lines correspond to the $C_+$ solution while the solid lines
represent the $C_-$ solution. 
The dotted vertical line represents the critical wavenumber calculated using
Eq.~(\ref{kxcrit}).  }\label{growth1} \end{figure}

\begin{figure}[!hh] \center{\includegraphics[width=8cm]{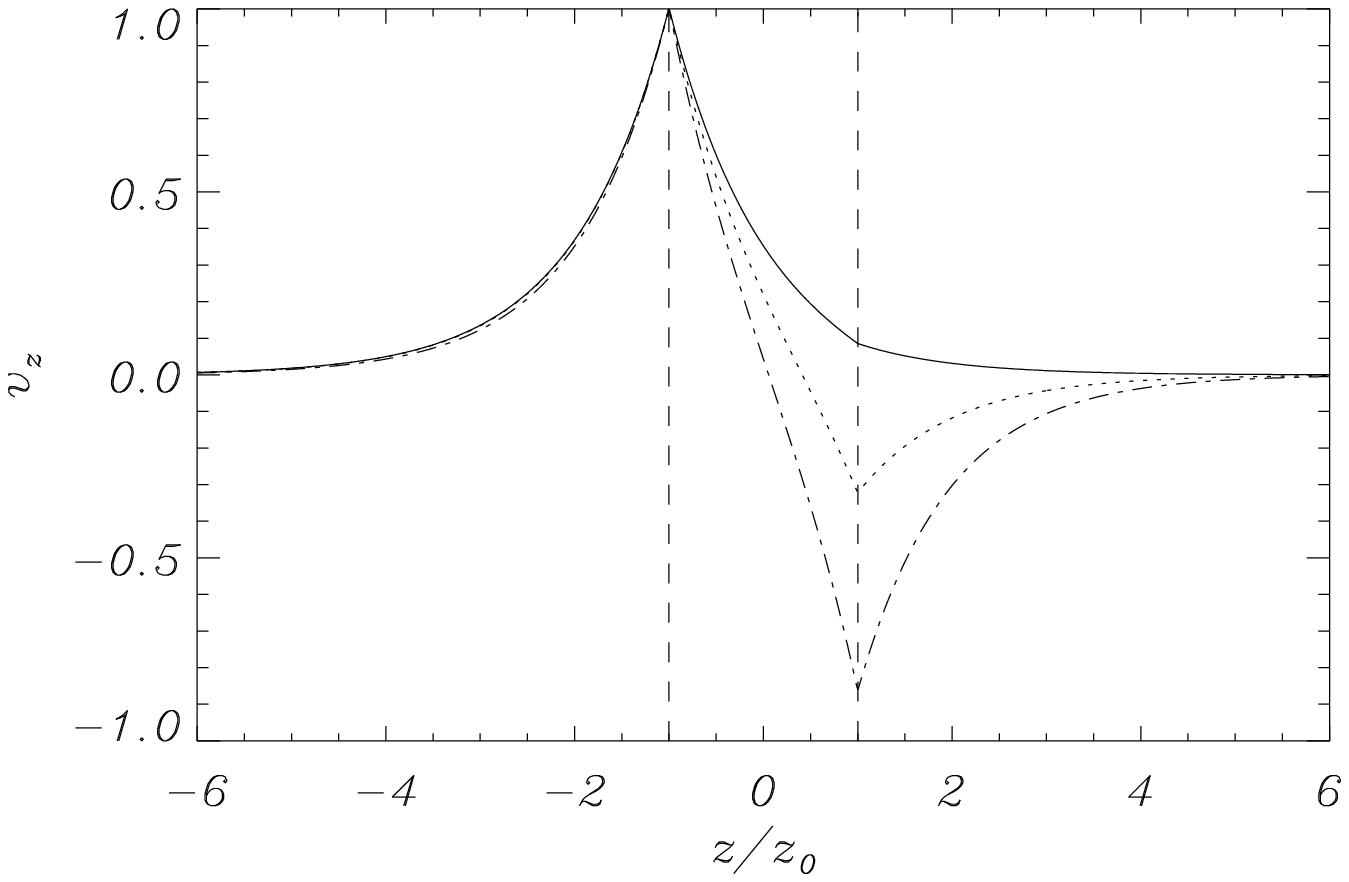}}
\center{\includegraphics[width=8cm]{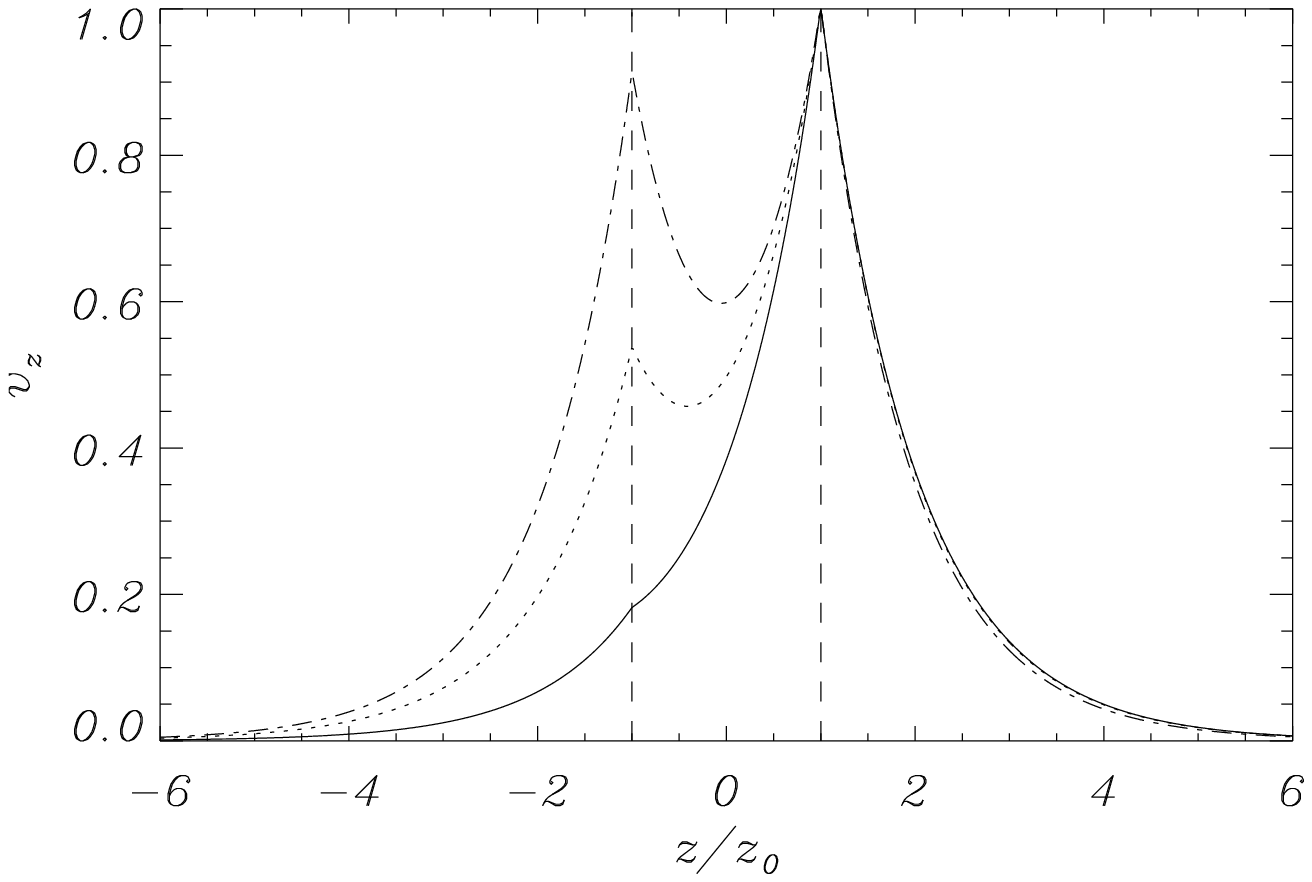}}  \caption{\small  Eigenfunctions
in the presence of gravity and magnetic field  associated to the $C_-$
solution (top panel) and the $C_+$ solution (bottom panel).  The continuous line
correspond s to $\lambda=2\times 10^4 \, \rm km$, the dotted line to
$\lambda=6\times 10^3 \, \rm km$, while for the dash-dotted curve
$\lambda=2\times 10^3 \, \rm km$. In this plot $R=10^2 \, \rm km$, $\rho_{\rm
p}=100 \rho_{\rm c}$ and $v_{\rm Ap}=100\,\rm km\, s^{-1}$. The dashed lines
represent the slab boundaries. }\label{eigen} \end{figure}

When gravitational and magnetic forces are present at the same time we expect to
find modes with mixed character. As in the interface problem, the interplay of
these two restoring forces might cause the unstable mode to become stable if the
magnetic field is strong enough, or if the wavelength is larger than a critical
value. This behaviour is associated to the $C_-$ solution in our notation. On
the contrary, the stable mode found in the situation with gravity only, remains
always stable when the magnetic field is different from zero. In this case the
solution corresponds to the $C_+$ constant. An example of this behaviour is
found  in Figure \ref{growth1}, where the square of the frequency calculated
using the full dispersion relation given by Eq.~(\ref{instslab}), is plotted as
a function of the wavenumber. We identify the two modes, the stable solution
(dashed line) and the initially unstable mode (continuous line) that becomes
stable when the wavenumber is larger than a critical value. We have used the
same choice of parameters, meaning that we are in the regime  $k_x\ll k_y$.  The
solutions associated to the individual interfaces, calculated using
Eq.~(\ref{instinter}), are also plotted in Figure \ref{growth1} with thin lines.
The agreement between the slab and interface solutions is quite good for small
$k_x$, meaning that for the particular parameters we are using, the modes are,
to all practical purposes, surface waves associated to the individual interfaces
in this regime. Thus, for $k_x$ small there are no global kink-like transverse
modes. Further confirmation of this behaviour is found in the eigenfunctions,
represented in Figure \ref{eigen} for $v_{\rm Ap}=100\,\rm km\, s^{-1}$ and
$\lambda=2\times 10^4 \, \rm km$ (see continuous lines). The surface waves are
barely affected by the presence of the other interface, and the overall
structure is quite similar to the one found in the non-magnetic case (compare
with Figure \ref{vzeigenb0}).

The weakly coupled behaviour of the two interfaces manifested in
Figure \ref{eigen} is better understood by
analysing the value of $C$ in the dispersion relation. If we assume that $\rho_{\rm c}\ll
\rho_{\rm p}$ the terms proportional to $\rho_{\rm c}/\rho_{\rm p}$ can be
neglected in Eqs.~(\ref{eeq}) and (\ref{feq}). It is easy to see that with this
simplification, and when the following condition is
satisfied
\begin{eqnarray}\label{cond}
gk \sinh{2kz_0}\gg  k_x^2 v_{\rm Ap}^2,
\end{eqnarray}
the constants $E$ and $F$ tend to same value. This means that for the unstable
mode the constant $C_-$ goes to zero, and thus for this mode the slab dispersion
relation tends to the interface dispersion relation given by
Eq.~(\ref{instintersim}). The equivalent result is found for the stable mode.  Therefore,
if the condition given by Eq.~(\ref{cond}) is satisfied, the system is well
described by the interface results. Note that Eq.~(\ref{cond}) does not mean
that the role of the magnetic field is neglected in front of the gravitational
term, the interface dispersion relation still contains the magnetic term. In this regime we can make further
analytical progress assuming that $k_x\ll k_y$, and reducing
Eq.~(\ref{instintersim}), for the unstable mode, to
\begin{eqnarray}\label{instslabsimp}
\omega^2= -\frac{g}{R}+2 k_x^2  v_{\rm Ap}^2,
\end{eqnarray}
where we have used that $k_y=1/z_0=1/R$.

If we assume that the thread radius and the Alfv\'en  speed are known from
observations, we can determine the critical wavenumber required to have a stable
configuration ($\omega^2>0$)
\begin{eqnarray}\label{kxcrit}
k_x^c=\sqrt{\frac{1}{2}\frac{g}{R}} \frac{1}{v_{\rm Ap}}. \end{eqnarray}
In Figure
\ref{growth1} this critical wavenumber is represented by a vertical dotted
line. Alternatively, if the wavenumber is known the Alfv\'en speed required to have a
stable configuration is 
\begin{eqnarray}\label{vacrit} v_{\rm
Ap}^c=\sqrt{\frac{1}{2}\frac{g}{R}} \frac{1}{k_x}. \end{eqnarray}
This
equation might be useful for some applications, as we will show in section
\ref{app}.

The condition given by Eq.~(\ref{cond}) is not fully satisfied if, for example,
we chose a wavelength of the order of $\lambda \sim 6\times 10^3 \, \rm km$, and
we consider that the Alfv\'en speed is $v_{\rm
Ap}=100\,\rm km\, s^{-1}$, i.e. when we move to the right in the dispersion
diagram shown in Figure \ref{growth1}. For these parameters the modes are no
longer pure surface waves associated to the interface and have a mixed nature,
as can be appreciated in Figure \ref{eigen} (see dotted lines). In fact in this figure the particular
choice of parameters satisfies that 
\begin{eqnarray}\label{condcoupling} gk
\sinh{2kz_0}\approx  k_x^2 v_{\rm Ap}^2. \end{eqnarray} 

\noindent This condition means that the coupling between gravitational and magnetic
forces is strong,  and the resulting modes a mixture between interface
and slab waves. The situation is intermediate between the purely gravitational
and the purely magnetic modes. Finally, when 
\begin{eqnarray}\label{condcoupling1}
gk \sinh{2kz_0}\ll k_x^2 v_{\rm Ap}^2,
\end{eqnarray}
the modes have a global nature and are very similar to the
sausage and kink modes (see Figure
\ref{eigen}, dash-dotted lines) since they are dominated by magnetic forces. This case is almost
equivalent to the purely magnetic situation discussed in subsection
\ref{magnsec},  and the modes are always stable.


Figure \ref{eigen} shows very clearly the transformation from single surface
waves to global sausage and kink modes when the Alfv\'en speed is fixed and the
wavelength is allowed to change, but a completely equivalent behaviour is found
if the wavelength is fixed and the Alfv\'en speed varies. Note also that
according to the dispersion diagram given by Figure~\ref{growth1} we find that
the solution associated to the $C_-$ constant changes from unstable to stable,
and the eigenfunction evolves from an interface solution associated to the lower
interface in the gravitationally dominated case to a sausage mode in the
magnetically dominated situation. On the other hand, the solution associated to
the upper interface, the $C_+$ mode, is always stable and it eventually tends
the global kink mode.

\subsection{Thread lifetimes}

From the previous results we conclude that in case of instability it will be
triggered at the bottom of the thread and it is due to the $C_-$ solution. The development of this instability will
eventually produce significant changes in the whole thread structure. Here we
want to calculate the time required to observe the effects on the upper
interface. We know that the velocity of the lower interface is (in
the linear regime)
\begin{eqnarray}\label{vel}
v=v_0e^{t/\tau}, 
\end{eqnarray}
where $\tau=1/|\omega|$ is the growth rate of the instability. If the interface moves upward then the vertical displacement is
\begin{eqnarray}\label{displ}
\xi=v_0 \tau e^{t/\tau}+\xi_0.
\end{eqnarray}
We assume that at $t=0$ the interface is located at $z=-z_0$ (according to our slab
model),
this means that
\begin{eqnarray}\label{disp}
\xi_0=-\left(z_0+v_0 \tau\right).
\end{eqnarray}
\noindent Now using  Eq.~(\ref{displ}) we can calculate the time required to have a displacement of
the interface from $-z_0$ to $z_0$
\begin{eqnarray}\label{tmix}
t_{\rm m}=\tau \ln{\left(\frac{2 z_0}{v_0 \tau}+1\right)}.
\end{eqnarray}
This parameter provides a rough measure of the time to blend the lower and 
upper interfaces of the thread and, as a result of this process there is a
mixture of the thread density with the much lower coronal density. The time
scale $t_{\rm m}$ can be interpreted as the disappearance time or lifetime of
the thread due to the Rayleigh-Taylor instability. However, we have to bear in
mind that the previous calculation is based on the assumption of a linear
regime, but in a real situation the nonlinear effects might increase the
lifetimes due to the saturation of the instability.




\section{Application to observations of oscillating threads}\label{app}

\cite{linetal07} found travelling waves in a quiescent filament from the
analysis of  velocity Dopplergrams. The observed threads have a radius of the
order of $100\, \rm km$, and their typical lifetime is $9\, \rm min$. For the
particular thread marked with the label L2 in \cite{linetal07}, the period is
$5.4\, \rm min$, the estimated wavelength of the propagating waves is $3000\,
\rm km$, and the phase speed is around $9.5\, \rm km\, s^{-1}$. The amplitude of
the oscillations is $1-2\, \rm km\, s^{-1}$. 

\begin{figure}[!hh] \center{\includegraphics[width=8cm]{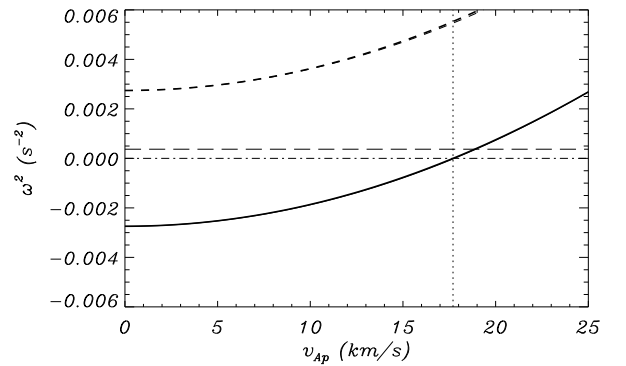}} 
\caption{\small Square of the frequency as a function of the Alfv\'en speed in
the thread. In this plot $R=10^2 \, \rm km$, $\lambda=3 \times 10^3 \, \rm km$ 
and $v_{\rm Ap}^c\approx 17.7\, \rm km\, s^{-1}$. The same notation as in the
previous figures has been used. The long dashed line corresponds to the
frequency ($P=5.4\, \rm min$) derived from the observations of
\cite{linetal07}. The dashed lines correspond to the $C_+$ solution while the solid lines
represent the $C_-$ solution.}\label{growth4} \end{figure}

We concentrate first on the travelling wave. Since the wave has a propagating
character we interpret this mode, according to our model, as a stable solution,
the unstable mode simply grows with time and it does not propagate along the
thread producing oscillations (in the absence of longitudinal flows). The thread radius and the wavelength of the
perturbation are known from observations, therefore we calculate the critical
Alfv\'en speed to have stable solutions. From Eq.~(\ref{vacrit}) it is found
that $v_{\rm Ap}^c\approx 17.7\, \rm km\, s^{-1}$. This is the lower bound to
have stable modes whose wavelength is equal to or smaller than $3000\, \rm km$. The
Alfv\'en speed in the thread might be higher than  $17.7\, \rm km\, s^{-1}$, and
the exact value  is calculated using the period of oscillation known from
observations, and the dispersion relation. From the two possible modes, only the
solution associated to the lower interface can be matched with the parameters
determined from observations. Figure \ref{growth4} shows the dependence of the
square of the frequency with the Alfv\'en speed in the thread. The Alfv\'en
speed that produces a match with the period of $5.4\, \rm min$ (and the
corresponding phase speed of $9.5\, \rm km\, s^{-1}$) is $18.8\, \rm km\,
s^{-1}$, and has been calculated using Eq.~(\ref{instslabsimp}).  This value
corresponds to the intersection of the long dashed line, representing the
reported period, and the thick continuous line associated to the dispersion
relation. Note that the difference between the critical and the estimated
Alfv\'en speed in the thread is rather small. 

\begin{figure}[!hh] \center{\includegraphics[width=8cm]{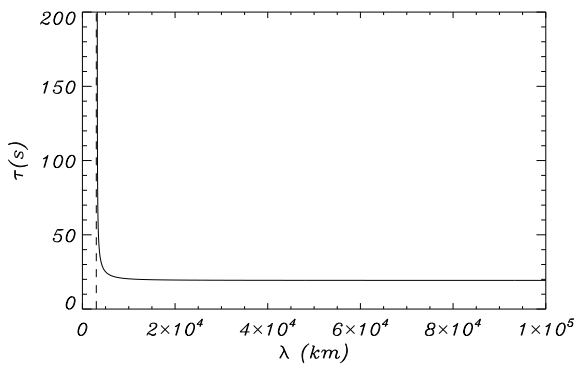}} 
\caption{\small Growth rate ($\tau=1/|\omega|$) as a function of the wavelength
of the perturbation. In this plot $R=10^2 \, \rm km$ and $v_{\rm Ap}= 18.8\, \rm
km\, s^{-1}$. The dashed line denotes the wavelength, $\lambda_c= 3.2\times
10^3\, \rm km$, where the mode changes from stable to unstable.}\label{growthlamb} \end{figure}

Once we have an estimation of the Alfv\'en speed in the thread we calculate the
critical wavelength to have a stable situation. Using Eq.~(\ref{kxcrit}) we
obtain a value of $3200\, \rm km$. This value is slightly larger than the
wavelength of the propagating wave, $3000\, \rm km$. In Figure \ref{growthlamb}
the growth rate is plotted as a function of the wavelength, being the lower
value the critical wavelength. Close to the critical wavelength the growth times
can be very large but for $\lambda\geq 10^4\, \rm km$ the growth rate is almost
constant with a value of $20\, \rm s$. Using this growth rate, the value of
$z_0$, and the value of the amplitude $v_0$ assumed to be of the order of $1\,
\rm km\, s^{-1}$, we estimate from Eq.~(\ref{tmix}) that the mixing time is
$t_{\rm m}=48\, \rm s$. This value is one order of magnitude smaller than the
estimated lifetime of the thread, which is around $540\, \rm s$ ($9\, \rm min$).
In this particular case the instability is too efficient and according to our
model this thread should disappear in less than one minute.

\begin{figure}[!hh] \center{\includegraphics[width=8cm]{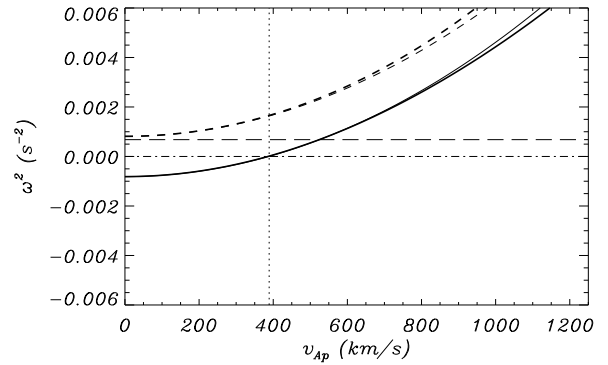}} 
\caption{\small Square of the frequency as a function of the Alfv\'en speed in
the thread. In this plot $R=3.3\times 10^2 \, \rm km$, $\lambda=1.2 \times
10^5\, \rm km$ and $v_{\rm Ap}^c\approx 389.1\, \rm km\, s^{-1}$. The same
notation as in the previous figures has been used. The long dashed line
corresponds to the frequency ($P=4\, \rm min$) derived from the observations of
\cite{okamoto07}.  The dashed lines correspond to the $C_+$ solution while
the solid lines represent the $C_-$ solution.}\label{growth5} \end{figure}

Another example of thread oscillations has been investigated by
\citet{okamoto07}.  Using $Hinode$ observations  these authors found transverse
motions of threads in an active region prominence. \citet{terrarr08} performed a
seismological analysis of the same event. The radius of the threads are in the
range $180-330\, \rm km$ and their length is $1700-16,000\, \rm km$. The
estimated minimum length of the magnetic tube is $L=10^5\, \rm km$. We
concentrate on the thread of length $16,000\, \rm km$ since this structure has
the smallest longitudinal motion (the horizontal flow velocity is around $15\,
\rm km\,s^{-1}$). Contrary to the situation studied by \cite{linetal07} the
oscillations seem to have a standing nature, and the thread is moving vertically
with a period of $4\,\rm min$. Another difference with respect to
\cite{linetal07} is that we have information about the maximum wavelength that
satisfies the line-tying conditions. It corresponds to the fundamental mode
($\lambda=2L$) meaning that for a fully filled thread $\lambda=2\times 10^5\, \rm
km$. However, we can take into account the fact that the tube is only partially
filled. \citet{soleretal10} calculated the frequency of oscillation for a thread
of length $L_{\rm p}$ (the total length of the tube is $L$). From the comparison
of this frequency with the frequency of oscillation of the fully filled thread
they found that the effective wavelength that should be used is $\lambda=\pi
\sqrt{\left(L-L_{\rm p} \right) L_{\rm p}}$. This expression is valid as long as
$L_{\rm p}/L<0.4$. For the thread we are considering we have $L_{\rm p}=16,000\,
\rm km$ while $L=10^5\, \rm km$, ($L_{\rm p}/L=0.16$), and the corresponding
effective wavelength is $\lambda=1.2\times 10^5\, \rm km$. This is a simple way
to mimic the effect of a partially filled tube. Using this effective value for
the wavelength we calculate from Eq.~(\ref{vacrit}) that the critical Alfv\'en
speed to have a stable situation is $389.1\, \rm km\, s^{-1}$. Therefore, if the
Alfv\'en speed is larger than this value the thread is going to be always stable
with respect to the Rayleigh-Taylor instability. Figure \ref{growth5} shows the
dependence of the square of the frequency with the Alfv\'en speed in the thread
for the previous example. Again, as for the situation studied in \cite{linetal07}
only the mode associated to the lower interface can be fitted with the reported
period, and the inferred Alfv\'en speed in the thread is $525.8\, \rm
km\,s^{-1}$. This speed is in the range of values estimated by \citet{terrarr08}.
However, the interpretation in terms of the interface mode does not agree with
observations since for this particular example the whole thread shows a clear
kink-like vertical motion. Again the effect of gravity seems to be too strong in
the slab model.

\section{Discussion and conclusions}

\begin{figure}[!hh] \center{\includegraphics[width=8cm]{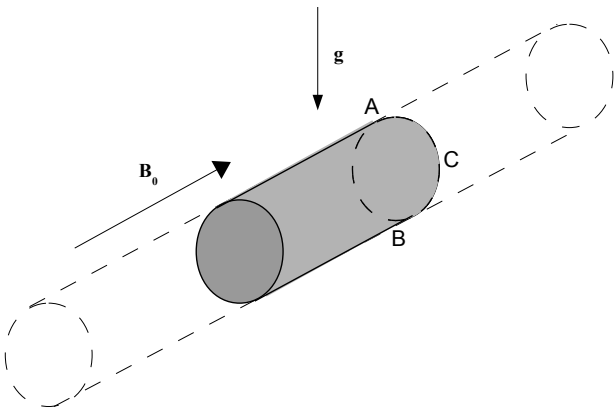}}
\caption{\small Sketch of the three-dimensional cylindrical  model representing
a prominence thread. The projection of the gravity force varies along the
circular interface.}\label{fibril3d} \end{figure}

In this work threads have been modelled as Cartesian slabs. The presence of
gravity acting perpendicularly to the axis of the structure does not allow us to
have a simple equilibrium using a more appropriate geometry. The most elemental
models in cylindrical geometry involve the presence of magnetic twist \citep[see
for example][]{lerchelow80,lowzhang04,petrieetal07,bloklandkeppens11a}. To
simplify things and to have an analytical treatment of the problem, instead of a
thread with cylindrical cross-section we have adopted the slab configuration. We
have supposed that the slab is infinite in the ignorable direction, and more
important, we have assumed that there is dense material along the whole tube
since the gravity force prevents to have a simple analytical equilibrium between
the dense and the evacuated part of the thread along the axial magnetic field.
These features of our model certainly affect the frequencies of oscillation of
the stable modes and the growth rates of the unstable modes. In addition, the
effect of gravity is overestimated in the slab configuration because the gravity
force is perpendicular to all the points on the interfaces separating the two
fluids (see Figure \ref{fibril}). In a more realistic three-dimensional model,
such as a cylindrical tube, the overall contribution of the gravity force
perpendicular to the interface is weaker. This can be understood from Figure
\ref{fibril3d}, gravity is perpendicular to the interface only at points $A$ and
$B$, while at point $C$ the projection across the interface is zero. 
However, we have to bear in mind that a realistic geometry could be anything
between a slab and a flattened cylinder, which might be closer to a slab of
finite width rather than to a pure circular cross-section cylinder.

We have shown that the mode that fits with the frequency reported
in the observations is always associated, according to our model, to the lower
interface and does not have a global nature. This is in contradiction with the
observed motions of \citet{okamoto07} displaying a global vertical displacement
of the whole thread. If the effect of gravity on the eigenfrequencies is
smaller, as we expect for a more realistic configuration, then a match in
frequency for the global mode is possible (the upper curve in Figure
\ref{growth5} might intersect the dashed line). Additionally, the growth rate of
the unstable mode will increase, which is one of the questionable results of the
slab model since the growth of the instability is too fast, according to  the
comparison with the observations of \cite{linetal07}.  

We have also adopted in our analysis the incompressible assumption for the
perturbations. It is known that, in general, the effect of compressibility helps
to stabilise the unstable modes. Since gas pressure must change with position to
have hydrostatic equilibrium, the analytical investigation  of the compressible
case is much more complicated than the incompressible situation due to the
presence on non-constant coefficients in the equations (for example, the sound
speed would change across the slab). However, even under the compressible
assumption, in the regime we have considered in this work, $k_x \ll k_y$, 
motions are essentially incompressible. A clear example of this situation is the
kink mode in the compressible slab \citep[see][]{arrterretal07} or the kink
mode in cylindrical geometry in the thin tube limit
\citep[see][]{goossensetal09}. Hence, incompressibility is a reasonable
approximation that has the additional advantage that the problem can be studied
fully analytically, this has allowed us to  derive the algebraic dispersion
relation given by Eq.~(\ref{instslab}). Notice also that the choice of a step
function for the density is convenient to avoid a situation where the Alfv\'en
speed smoothly changes with height, since this would imply the existence of
resonances for the case with $k_y\neq 0$.  

From the analysis of the dispersion relation and the eigenfunctions we have shown
that, contrary to the purely magnetic case, the global transverse kink mode  does
not necessarily exist for the whole range of parameters in the stratified model,
even in the stable regime. This might have some important consequences, since
other types of motions associated to the individual interfaces, or of mixed
nature between surface and global kink waves, are feasible according to our
model. Thus, from the observational point of view it might be interesting to try
to identify  such kind of motions since their detection can give more clues about
the real physical conditions in threads, specially about the interplay between
magnetic and gravitational forces.

Using the slab model we have demonstrated that the system allows a mode that is
always stable, independently of the value of the Alfv\'en speed in the thread,
while there is another mode that is Rayleigh-Taylor unstable but becoming stable
when the Alfv\'en speed is increased. Hence, gravity might have a strong effect on
the modes of oscillation of horizontal threads, depending on the strength of the 
magnetic field. In the case of threads in quiescent filaments like the one studied
by \cite{linetal07}, where the Alfv\'en speed presumably low, too short lifetimes
are found using the slab model, i.e., the instability is too efficient. On the
contrary, in active region prominences \citep[like in][]{okamoto07}, the
stabilising effect of the magnetic tension might be enough to suppress the
Rayleigh-Taylor instability for a wide range of wavelengths. We have also shown
that the instability can  be used to establish the minimum Alfv\'en speed in the
thread, an approach that is different from other seismological analysis of thread
oscillations \citep[see for example][]{terrarr08,linetal09,soleretal10}.
Nevertheless, more elaborated models need to be explored in detail before a 
reliable comparison of periods and thread lifetimes with observations can be
performed.

Finally, we have to remark that the classification of the modes and their
spatial structure strongly depends on the slab configuration. As we have already
mentioned, the sausage mode found in our system when the magnetic field is
dominant does not exist as a trapped mode in the cylindrical tube. This mode is
precisely becoming unstable when gravity dominates over the magnetic field. This
means that the geometry can have an important effect on the spectrum of the
modes, and a careful analysis using more realistic geometries \citep[see][
for flattened cylindrical models]{bloklandkeppens11} is required to understand
well the effect of gravity in horizontal threads.







\acknowledgements J.T. acknowledges support from the Spanish Ministerio de
Educaci\'on y Ciencia through a Ram\'on y Cajal grant. All the authors
acknowledge the funding provided under the project AYA2011-22846 by the Spanish
MICINN and FEDER Funds.  The financial support from CAIB through the ``Grups
Competitius'' scheme is also acknowledged. The authors also thank I. Arregui, A.
J. D{\'{\i}}az and R. Soler for their comments and suggestions that helped to
improve the paper.


\end{document}